\documentclass[aps,pra,twocolumn,amsmath,amssymb,superscriptaddress,notitlepage,groupedaddress,longbibliography]{revtex4-1}
\usepackage{graphicx}
\usepackage{times,psfrag,subfigure}
\usepackage{enumerate}
\usepackage{dsfont}
\usepackage{dcolumn}
\usepackage{bm,bbm}
\usepackage{hyperref}
\usepackage{color}
\usepackage{latexsym,amsmath,amssymb,bm,euscript,ulem}
\usepackage{textcomp}
\usepackage{tabularx}
\usepackage{setspace}
\usepackage{ctable}
\usepackage{sidecap}
\usepackage{placeins}
\usepackage{threeparttable}
\usepackage{multirow}
\usepackage{pifont}

\let \oldbm \bm
\renewcommand{\vec}[1]{\oldbm{#1}}

\def\bk{{\vec k}}

\def\bA{{\vec A}}

\def\bq{{\vec q}}

\def\bG{{\vec G}}

\def\bm{{\vec m}}

\def\br{{\vec r}}

\newcommand{\inner}[2]{\langle #1|#2\rangle}

\def\tr{\mathop{\mathrm{tr}}}

\def\T{\mathcal{T}}

\def\P{\mathcal{P}}

\def\H{\mathcal{H}}

\def\M{\mathcal{M}}

\newcommand{\beq}{\begin{equation}}
\newcommand{\eeq}{\end{equation}}
\newcommand{\beqarray}{\begin{eqnarray}}
\newcommand{\eeqarray}{\end{eqnarray}}

\begin{document}

\title{Nematic topological semimetal and insulator in magic angle bilayer graphene at charge neutrality}

\author{Shang Liu}
\affiliation{Department of Physics, Harvard University, Cambridge, MA 02138}

\author{Eslam Khalaf}
\affiliation{Department of Physics, Harvard University, Cambridge, MA 02138}

\author{Jong Yeon Lee}
\affiliation{Department of Physics, Harvard University, Cambridge, MA 02138}

\author{Ashvin Vishwanath}
\affiliation{Department of Physics, Harvard University, Cambridge, MA 02138}

\date{\today}

\begin{abstract}
We report on a fully self-consistent Hartree-Fock calculation of interaction effects on the Moir\'e flat bands of twisted bilayer graphene, assuming that valley U(1) symmetry is respected. We use realistic band structures and interactions and focus on the charge neutrality point, where experiments have variously reported either insulating or semimetallic behavior. Restricting the search to orders for which the valley U(1) symmetry remains unbroken, we find three types of self-consistent solutions with competitive ground state energy (i) insulators that break $C_2 {\mathcal T}$ symmetry, including valley Chern insulators (ii) spin or valley polarized insulators and (iii) rotation $C_3$ symmetry breaking semimetals whose gaplessness is protected by the topology of the Moir\'e flat bands. We find that the relative stability of these states can be tuned by weak strains that break $C_3$ rotation. The nematic semimetal and also, somewhat unexpectedly, the $C_2 {\mathcal T}$ breaking insulators, are stabilized by weak strain. These ground states may be related to the semi-metallic and insulating behaviors seen at charge neutrality, and the sample variability of their observation. We also compare with the results of STM measurements near charge neutrality. 
\end{abstract}

\maketitle

\section{Introduction}
The discovery of interaction-driven insulating and superconducting behavior in twisted bilayer graphene (TBG) \cite{PabloMott, PabloSC} has inspired intensive efforts to understand this behavior \cite{Balents18, Po2018, IsobeFu, Thomson18, YouAV, Vafek, Xie2018,Nandkishore, Kivelson, PhilipPhillips,phononMacDonald, phononLianBernevig, Zou2018, Bitan19} and to find related systems which exhibit similar phenomenology \cite{Zhang2018, Khalaf2019, Mora19, Cea19, Bi2019Strain, Lee19}. This work has started to bear fruit with several groups announcing similar observations in TBG samples \cite{Dean-Young, Sharpe2019, Efetov, OhioBelow} as well as other Moir\'e materials \cite{FengWang-TLGonBN-SC, FengWang-TLG-Mott,TDBGexp2019,IOP_TDBG,PabloTDBG}. The basic mechanism underlying the enhancement of correlation in these materials is understood to originate from the long-wavelength Moir\'e pattern leading to quenching of the electron kinetic energy manifested in flat energy bands \cite{MacDonald2011, Santos}. Nevertheless, the nature of the observed correlated insulating states remains under debate \cite{Po2018, Thomson18, IsobeFu, Vafek, Xie2018}. 

Early experiments found clear signature of a correlated insulating state at half-filling \footnote{We follow the standard convention of measuring the filling relative to charge neutrality where the Dirac points of the original graphene sheet reside. Complete filling corresponds to $\nu_T=4$ while completely empty $\nu_T=-4$ electrons per Moir\'e unit cell, accounting for both spin and valley degeneracy. Half filling corresponds to $\nu_T=\pm 2$}  \cite{PabloMott, Dean-Young}. 
Subsequently, insulating ferromagnetic states were also observed at quarter and three-quarter fillings \cite{Dean-Young, Sharpe2019}. On the other hand, in all these experiments\cite{PabloMott, PabloSC, Dean-Young} insulating behavior was absent at charge neutrality (CN) where signatures of semimetallic behavior were observed instead. In contrast, a recent experiment surprisingly found an insulator at CN with a transport gap exceeding those at 1/2, 1/4 and 3/4 fillings \cite{Efetov}. 

On the theory side, it was realized early on \cite{Po2018, po2018faithful} that a simple Mott picture for the insulating phase is complicated by the band topology which prohibits the construction of localized orbitals describing the flat bands while preserving all the symmetries. Various orders have been proposed to account for the insulating states.  At charge neutrality, a $C_2 \T$ symmetry breaking insulator, 
 with Chern number $\pm 1$ for each spin and valley flavour was proposed in \cite{Xie2018},  along with a $C_2 \T$ symmetry preserving insulator that is believed to require mixing with remote bands \cite{Xie2018}. An intervalley coherent order \cite{Po2018} was proposed as a candidate for the insulating state at half-filling while  nematic orders were  discussed in \cite{Kivelson,Thomson18,Sboychakov19} and ferromagnetic ordering was proposed in \cite{Vafek}. In the presence of explicit $C_2$ symmetry breaking induced by a substrate, valley or spin polarized insulator with valley resolved Chern numbers have been discussed  \cite{Zhang2018, Bultinck19}.


In this letter, we perform a self-consistent Hartree-Fock mean field analysis for the screened Coulomb interaction projected onto the flat bands.

We focus on the CN point because of its pivotal role in determining the entire phase diagram. We will discuss other fillings in subsequent work. For simplicity, we restrict our attention to orders that conserve the U(1) valley charge as well as translation invariance at the scale of the Moir\'e unit cell. Our results include the expected spin-polarized and valley-polarized insulators, which break no other symmetries. In addition we observe a strong tendency to breaking spatial rotation symmetries. We find a $C_2 \T$-breaking insulator and two distinct  $C_2 \T$-symmetric semimetallic phases which break $C_3$-symmetry 
.  The  $C_2 \T$-breaking phase has a Chern number of $C=\pm1$, per flavor (i.e. valley and spin). Different spin/valley orderings then lead to various ground states ranging from quantum anomalous Hall (QAH) to quantum valley Hall (QVH) or quantum spin Hall (QSH) insulators with very similar Hartree-Fock energies. Of these, the QVH insulator breaks $C_2$ symmetry, allowing for a direct coupling to the $C_2$-breaking hBN substrate, potentially favoring it in situations where the sample and substrate are aligned.  
The gapless $C_3$-breaking phases are obtained by bringing the two Dirac cones from the Moir\'e $K$ and $K'$ very close to the $\Gamma$ point. Instead of merging and opening a gap as one might normally expect, the Dirac points remain gapless since they carry the same chirality \cite{Po2018}, a consequence of descending from the Dirac points of graphene from the same valley for the two layers.  This topological protection prevents them from annihilating, resulting in a gapless semimetallic state. Thus, the metallic nature at CN in this scenario is intimately tied to the topological properties of the magic angle flat bands. 

We investigate the effect of small explicit $C_3$ symmetry breaking which can arise in real samples due to strain \cite{Bi2019Strain}, and show that it strongly influences the competition between different symmetry broken phases, favoring one of the $C_3$-breaking semimetallic phases.  Surprisingly, the  $C_2 \T$-breaking insulators also exhibits a strong susceptibility to $C_3$ symmetry breaking.
The energies of the insulating $C_2 \T$-breaking and the semimetallic $C_2 \T$-preserving states are  lowered compared to the spin/valley ferromagnets, and approach each other quickly as the value of the $C_3$-breaking parameter is increased. Our results suggest that these two states are candidate ground states in the presence of very small explicit $C_3$ symmetry breaking which is likely to exist in experiments.  
Further competition between these two phases is likely to be settled by small sample-dependent details, potentially explaining the realization of an insulator in some samples and a semimetal in other samples. 



\section{Problem setup}
The single-particle physics is described by the Bistritzer-MacDonald (BM) model \cite{Santos,MacDonald2011}, which employs a continuum approximation close to $K$ and $K'$ for a pair of graphene sheets rotated relative to each other by an angle $\theta$. The Hamiltonian for the $K$ valley is given by: 
\begin{multline}
\H_+= \sum_l \sum_{\bk} f^\dagger_{l}(\bk) h_{\bk}(l\theta/2) f_{l}(\bk) \\ + \left(\sum_{\bk}\sum_{i=1}^3 f^\dagger_{t} (\bk+\bq_i) T_i f_{b}(\bk) + h.c.\right). \end{multline}
Here, $l=t/b\simeq \pm 1$ is the layer index, and $f_l(\bk)$ is the $K$-valley electron originated from layer $l$. $h_{\bk}(\theta)$ is the monolayer graphene $K$-valley Hamiltonian with twist angle $\theta$ (see Appendix for details) and $\bq_1$ is defined as $K_b - K_t$ with $K_l$ denoting the $K$-vector of layer $l$. $\bq_2=O_3\bq_1$ is the counterclockwise $2\pi/3$ rotation of $\bq_1$, and $\bq_3=O_3\bq_2$. Finally, the interlayer coupling matrices are given by
\beq
T_j = 
	\begin{pmatrix}
	w_0 & w_1 \, e^{- (j-1) \frac{2\pi i}{3}}\\
	w_1 \, e^{ (j-1) \frac{2\pi i}{3}} & w_0
	\end{pmatrix},
\eeq
with $w_0$ and $w_1$ denoting intrasublattice and intersublattice hopping, respectively. Due to lattice relaxation effects, which shrink the AA regions relative to the AB regions, the value of $w_0$ at the magic angle is about 75\% of $w_1$ \cite{Koshino2017, Koshino2018}. Throughout this work, we will use the values $w_1 = 110$ meV and $w_0 = 82.5$ meV. 
Explicit $C_3$ symmetry breaking is implemented via the substitution $T_1 \rightarrow (1 + \beta) T_1$ \cite{ZhangPo}.

 To study possible correlated insulating states, we employ a momentum-space self-consistent Hartree-Fock mean field theory. The momentum-space description allows us to focus on the pair of flat bands at CN, thereby evading the difficulties associated with the real space Wannier obstruction rooted in the fragile topology of these bands \cite{PoFragile, Po2018, po2018faithful}.Restricting the analysis to the flat bands 
 is only justified in the limit when both the bandwidth and interaction strength are much smaller than the bandgap. For realistic interactions, the interaction strength is of the same order as the bandgap which means that symmetry-broken states involving remote bands cannot be ruled out \cite{Xie2018}. However, given the strong angle dependence and appearance of correlated states only near the magic angle \cite{PabloMott, PabloSC, Dean-Young}, it is likely that the relevant symmetry-broken phases experimentally originate mostly from the two flat bands which justifies our approximation. We note, however, that our approach cannot capture possible changes in the topology of the flat bands arising from mixing with remote bands which were argued to be relevant in Ref.~\cite{Xie2018}. 

The Hartree-Fock (HF) mean field theory is defined in terms of the projector
\beq
P_{\alpha, \beta}(\bk) = \langle c^\dagger_{\alpha}(\bk) c_{\beta}(\bk) \rangle, \quad P(\bk)^2 = P(\bk) = P(\bk)^\dagger,
\eeq
where $c_\alpha(\bk)$ is the annihilation operator for an electron at momentum $\bk$ and $\alpha=(n,\tau,s)$ is a combined index for band, valley and spin, respectively. For a gapped or semimetallic phase at CN, $P$ satisfies $\tr P(\bk) = 4$ for all $\bk$ points\footnote{For semimetals, this is true everywhere except for the gapless points where the projector is not defined}. 
 The HF mean field Hamiltonian has the form
\begin{multline}
    \H_{\rm MF} = \sum_\bk \{c(\bk)^\dagger [h_0(\bk) + h_{\rm HF}(P,\bk)] c(\bk) \\ - \frac{1}{2} \tr h_{\rm HF}(P,\bk) P^T(\bk) \}.
    \label{HMF}
\end{multline}
Here, $c(\bk)$ is a column vector in the index $\alpha$, $h_0(\bk)$ denotes the single particle Hamiltonian and $h_{\rm HF}(\bk)$ is given by
\begin{multline}
h_{\rm HF}(P,\bk) = \frac{1}{A} \sum_{\bG} V_\bG \Lambda_\bG(\bk) \sum_{\bk'} \tr P^T(\bk') \Lambda^\dagger_\bG(\bk')\\ -\frac{1}{A} \sum_{\bq} V_{\bq} \Lambda_{\bq}(\bk) P^T(\bk + \bq) \Lambda^\dagger_{\bq}(\bk),
\label{hMF}
\end{multline}
where $A$ is the total area. The $\bk'$, $\bG$, and $\bq$ summations range over the first Brillouin zone, reciprocal lattice vectors, and all momenta, respectively. $V_\bq$ is the interaction potential which we take to be a single-gate-screened Coulomb interaction $V_\bq= \frac{e^2}{2 \epsilon\epsilon_0 q}(1- e^{-2qd_s})$ with dielectric constant $\epsilon = 7$ and screening length equal to the gate distance $d_s \approx 40$ nm.

The first term in (\ref{hMF}) is the Hartree term while the second is the Fock term. The matrix $\Lambda_\bq(\bk)$ contains the form factors for the single-particle 
\beq
[\Lambda_\bq(\bk)]_{\alpha,\beta} =  \langle u_\alpha(\bk)|u_\beta(\bk+\bq) \rangle.
\label{Lambda}
\eeq
where $\alpha$ and $\beta$ denote a combined index for spin, valley and band. The HF self-consistent analysis starts by proposing an ansatz for the projector $P(\bk)$, then substituting in the Hamiltonian (\ref{HMF}) which is then used to compute the new projector. This procedure is iterated until convergence is achieved. 

One important subtlety in the HF approach is that the band structure depends on the filling even without symmetry breaking. This follows from the fact that the form factor matrix $\Lambda_\bq(\bk)$ is not diagonal in the band index for $\bq \neq 0$ since the Bloch wavefunctions $u_{\alpha,\bk}$ from different bands are not orthogonal at different momenta. In addition, the Hartree term contains a trace over all filled bands which also affects the dispersion of the empty bands. This means that the band structure obtained from the BM model is only valid at a specific filling which determines the references point for out analysis. At this point, it will be assumed that interaction effects are already included in the parameters of the effective model which can be obtained by fitting to {\it ab initio} calculations or comparing to STM data away from the magic angles \cite{CaltechSTM, RutgersSTM, ColombiaSTM}. A natural choice of the reference point, which we adopt throughout this letter, is the CN point. This means that the single particle Hamiltonian $h_0(\bk)$ is given by
\beq
h_0(\bk) = h_{\rm BM}(\bk) - h_{\rm HF}(P_0,\bk)
\eeq
where $h_{\rm BM}(\bk)$ is the BM Hamiltonian and $P_0$ is the projector corresponding to symmetry unbroken state obtained by filling the lower band of the BM Hamiltonian at CN. We note that in the different approach of Ref.~\cite{Xie2018} where multiple bands are included,  the reference point was taken in the limit of decoupled layers. In our complementary approach of projection onto the flat bands it is not possible to implement such a choice.

\begin{figure}
    \centering
    \includegraphics[width=0.9\linewidth]{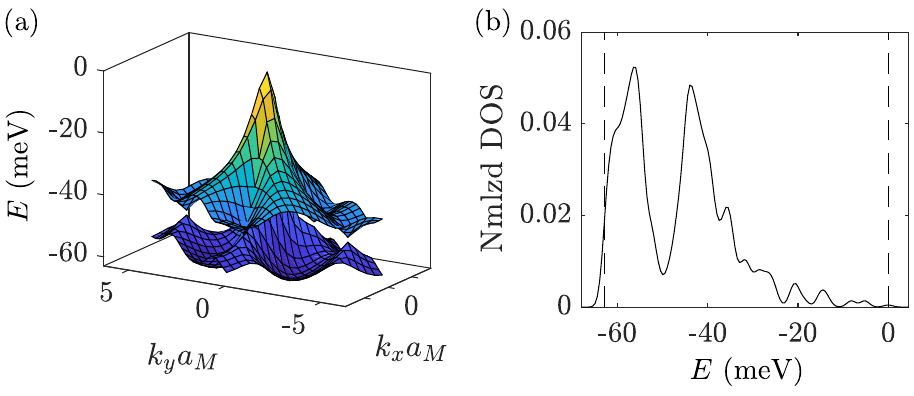}
    \caption{(a) $K$-valley band structure at empty filling. (b) Normalized density of states at empty filling, valid for both valleys. }
    \label{EmptyFilling}
\end{figure}

Using CN as our reference point implies that the bands at empty or full filling should include some HF corrections leading to a modified band structure shown in Fig.~\ref{EmptyFilling} together with the resulting density of states (DOS). We notice that the separation between the two peaks in the DOS is about 10-15 meV in agreement with the measured DOS in STM experiments \cite{CaltechSTM,CaltechSTM, RutgersSTM, PrincetonSTM}. Thus, our approach provides an explanation for the discrepancy between the experimentally measured peak separation and the expectation based on the BM model whose bandwidth close to the magic angle is {\em much} smaller (1-3 meV). 

\section{Symmetry-broken phases}
The interacting TBG Hamiltonian is characterized by the following symmetries \cite{Po2018}: spinless time-reversal $\T$ mapping the two valleys, {\rm SU}(2) spin rotation in each valley, {\rm U}(1) valley charge conservation and $C_6$ symmetry as well as a mirror symmetry which switches layers and sublattices but acts within each valley. Of these, only spin rotation, $C_3$, mirror  and $C_2 \T$ act within each valley. 
At integer filling, different correlated insulating phases can emerge by breaking some of these symmetries. Time-reversal symmetry is broken by valley polarized (VP) states, where the filling of the two valleys is different. Spin rotation symmetry is broken by spin polarization (SP) leading to ferromagnetic order. {\rm U}(1) valley charge conservation is broken in the presence of intervalley coherent (IVC) superposition of states from the two valleys. $C_2$ symmetry is broken by sublattice polarization which gaps out the Dirac points at the Moir\'e $K$ and $K'$ points.

Breaking $C_3$ symmetry alone does not generally lead to a gapped phase since it only moves the Dirac points away from the Moir\'e $K$ and $K'$ without gapping them out. Even strong $C_3$ breaking does not result in the merging and gapping of the Dirac nodes, in contrast with other familiar band structures such as single layer graphene. This follows from the fact that the chirality of the Dirac nodes, which is well defined in the presence of $C_2{\mathcal T}$ symmetry,  is the equal \cite{Po2018,Zou2018}, rather than opposite. Alternately, one can phrase the argument as follows. The topology of the two flat bands is captured by the second Stiefel-Whitney invariant $w_2$ \cite{StiefelWhitney,PoFragile,BernevigTBGTopology}. This invariant is protected by $C_2 \T$ and only depends on the flat band eigenstates which are unaffected by any symmetry breaking that does not involve other bands (which is the main assumption in this work).  Indeed, the $w_2$ invariant must be trivial  for a single isolated band, implying we cannot separate the pair of connected flat bands \footnote{In general, the total second Stiefel-Whitney number of two isolated bands is not equal to the sum of the second Stiefel-Whitney numbers separately for the two bands; there is an additional contribution from the first Stiefel-Whitney classes. In the special case we considered here, the $C_3$ symmetry of the two-band subspace guarantees this additional term vanishes.}. Thus the non-trivial $w_2$ invariant implies that the two Dirac points cannot be removed without breaking $C_2 \T$.


Before presenting the numerical results, let us make the following observations. First, it is relatively easy to show that a state with uniform full spin or valley polarization is always a self-consistent solution to the HF equations at CN for sufficiently narrow bands. These two states have the same energy in the absence of intervalley Hund's coupling \cite{Zhang2018, Lee19}.
The IVC state is known to have higher energy than SP/VP states for isolated  bands with non-vanishing valley Chern number \cite{Zhang2018, Bultinck19}. These arguments do not generalize to TBG where the extra symmetries of the problem complicate the discussion (see Appendix and Ref.~\cite{KIVC}). Nevertheless, we will exclude IVC orders from our numerical analysis  
 (which is equivalent to assuming unbroken U(1) valley charge conservation) since it leads to significant simplification by allowing us to focus on a single flavor (spin and valley). Different diagonal spin-valley orders can then be generated from the single-flavor solution by applying different symmetries. The energy competition with U(1) valley symmetry broken phases is considered in \cite{KIVC}.

\begin{figure}
	\centering
	\includegraphics[width=1\linewidth]{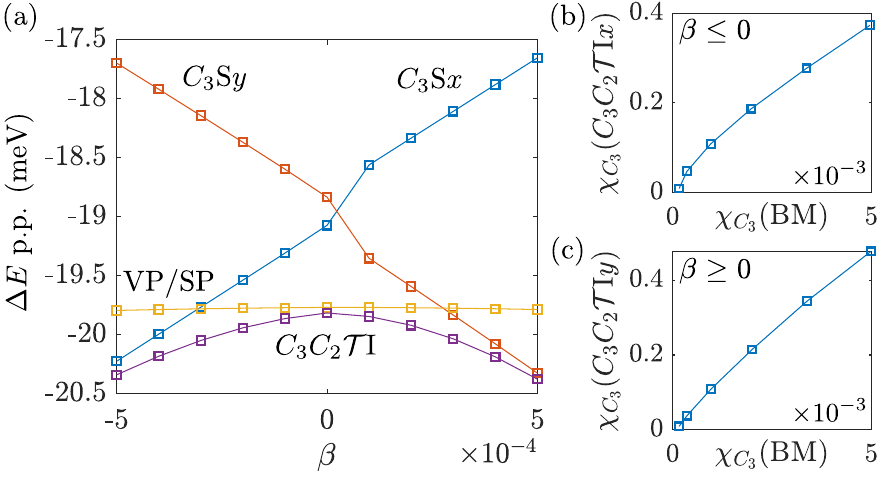}
	\caption{(a) Energies of the solutions of the self-consistent HF equations as a function of the $C_3$ symmetry breaking parameter $\beta$. All energies are measured relative to the state with no-broken symmetry. The degree of $C_3$-breaking measured by $\chi_{C_3}$ (Eq.~\ref{chi}) for the $C_2 \T$-breaking insulator as a function $\chi_{C_3}$ for the non-interacting system for $0 \leq |\beta| \leq 5 \times 10^{-4}$ are shown in panels (b) and (c) for positive and negative values of $\beta$, respectively.
	}
	\label{fig:EnStrain}
\end{figure}

\section{Results}
\begin{figure}[t]
	\centering
	\includegraphics[width=0.9\linewidth]{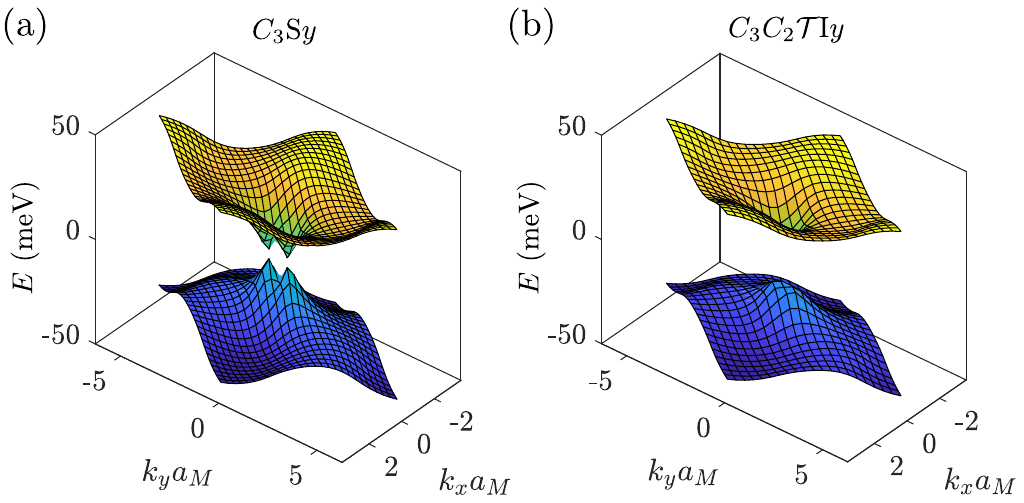}
	\caption{3D band structure for one of the (a) $C_3$-breaking semimetals and (b) $C_2 \T$ breaking valley Hall insulators.
	}
	\label{fig:BandStructure}
\end{figure}

The results for the self-consistent HF analysis are provided in Fig.~\ref{fig:EnStrain} showing the energies of the different solutions as a function of the $C_3$ symmetry breaking parameter $\beta$. There are two types of gapped solutions corresponding to either flavor-polarized (spin/valley) states or $C_2 \T$-breaking ($C_2 \T$I) insulators. There are three gapped solutions corresponding to spin-polarized (SP), valley-polarized (VP) and $C_2 \T$-breaking ($C_2 \T$I) insulators. 
The latter does not break $C_3$ when $\beta = 0$ but develops a large $C_3$ breaking component for $\beta \neq 0$. The extent of $C_3$ symmetry breaking can be quantified by defining 
\beq
\chi_{C_3} = \frac{1}{N} \sum_{\bk} (1 - |\langle \psi_{O_3 \bk}|C_3|\psi_{\bk} \rangle|^2),
\label{chi}
\eeq
which vanishes for any $C_3$-symmetric state. Here, $|\psi_{\bk} \rangle$ are the occupied single-particle eigenstates of the HF Hamiltonian with momentum $\bk$ (for a given flavor). The value of $\chi_{C_3}$ for the $C_2 \T$I state is shown in Fig.~\ref{fig:EnStrain} as a function of $\chi_{C_3}$ for the corresponding non-interacting states arising from explicit $C_3$-breaking parameter $\beta \neq 0$ in the BM model. We can clearly see from the figure that a relatively small $\chi_{C_3}(BM) \sim 10^{-3}$ in the non-interacting states induces induces a much larger $C_3$ symmetry breaking of almost two orders of magnitude in the $C_2 \T$I state. This serves to show that the $C_2 \T$I has very large susceptibility to $C_3$ symmetry breaking. 
In the following, we will refer to this state for positive and negative $\beta$ as $C_3 C_2 \T$I$y$ and $C_3 C_2 \T$I$x$, respectively. In addition to this insulating state, there are two distinct $C_2$-preserving semimetallic phases which spontaneously break $C_3$ even for $\beta = 0$ which we denote by $C_3$S$x$ and $C_3$S$y$, since they have Dirac points along $k_x$ and $k_y$, respectively. We notice that these ground states are similar to the ones obtained within a 10-band model Ref.~\cite{CaltechSTM} which used a  site-local  ansatz for the interactions. Examples of the band structures for the $C_3$S$y$ and $C_3 C_2 \mathcal{T}$I$y$ states are shown in Fig.~\ref{fig:BandStructure}.

 The $C_2\T$I phase obtained here is characterized by a Chern number of $\pm 1$ for a given spin and valley. The nature of the resulting phase depends on the precise symmetries which are broken as follows:  (i) if $\T$ is broken but $C_2$ and spin rotation are preserved, we obtain a quantum anomalous Hall (QAH) insulator with  total Chern number $C=\pm 4$, (ii) if spin rotation is broken we obtain a quantum-spin Hall (QSH) insulator with opposite Chern number $C_s=\pm 2$ for opposite spins, (iii) if $C_2$ is broken but $\T$ and spin rotation symmetry are preserved, then a quantum valley Hall (QVH) insulator obtains with opposite Chern number $C_v=\pm 2$ for opposite valleys, or (iv) we can additionally break spin rotation symmetry to obtain a state where opposite spins within the same valley also have opposite Chern numbers resulting in a quantum spin-valley Hall (QSVH) insulator where flipping either spin or valley flips the Chern number. There can also be states with total Chern number $\pm 2$, but these do not lead to any new physics or a lower energy, therefore we restrict to the above four possibilities for simplicity. 
 The QH, QSH and QVSH preserve either $C_2$ or a combination of $C_2$ and some internal symmetries, thus the only state expected to couple to the $C_2$-breaking hBN substrate is the QVH which is likely to be energetically favored in aligned samples\footnote{We thank A. Thomson for correspondence on this point}. The energies of the four possible $C_2 \T$ states are very similar differing only by very small $\sim 10^{-3}$ meV Hartree corrections. Similarly, the flavor-polarized state can be spin-polarized, breaking SU(2) spin rotation, valley-polarized breaking $\T$ or a spin-valley-locked states which breaks both but preserves a combination of $\T$ and $\pi$-spin rotation. These states are degenerate on the mean field level and are only distinguished by intervalley Hund's coupling \cite{Zhang2018, Lee19} neglected in this study. 

At $\beta = 0$, the energies of the insulating VP/SP and $C_2 \T$I states are very close and smaller by about 1 meV per particle than the energies of the two semimetallic $C_3$ breaking states. An explanation for this fact is provided in the next section in the limit where the intrasublattice hopping is switched off \cite{Tarnopolsky}. In this limit, we can establish a rigorous bound for the Fock energy which plays the dominant role in the energy competition. We will show that the SP/VP and the $C_2 \T$I states saturate the bound in this limit and should thus be approximately degenerate. The energies of the $C_3$S states are higher than these two. 
This analysis explains why these states are expected to be close in energy, but it is generally insufficient to capture the details of the energy competition which depends sensitively on the intrasublattice hopping $w_0$ and can only be determined numerically. A more detailed analysis of the effects of dispersion and finite sublattice symmetry breaking $w_0$ is given in Ref.~\cite{KIVC}.

\begin{figure}[t]
	\centering
	\includegraphics[width=0.9\linewidth]{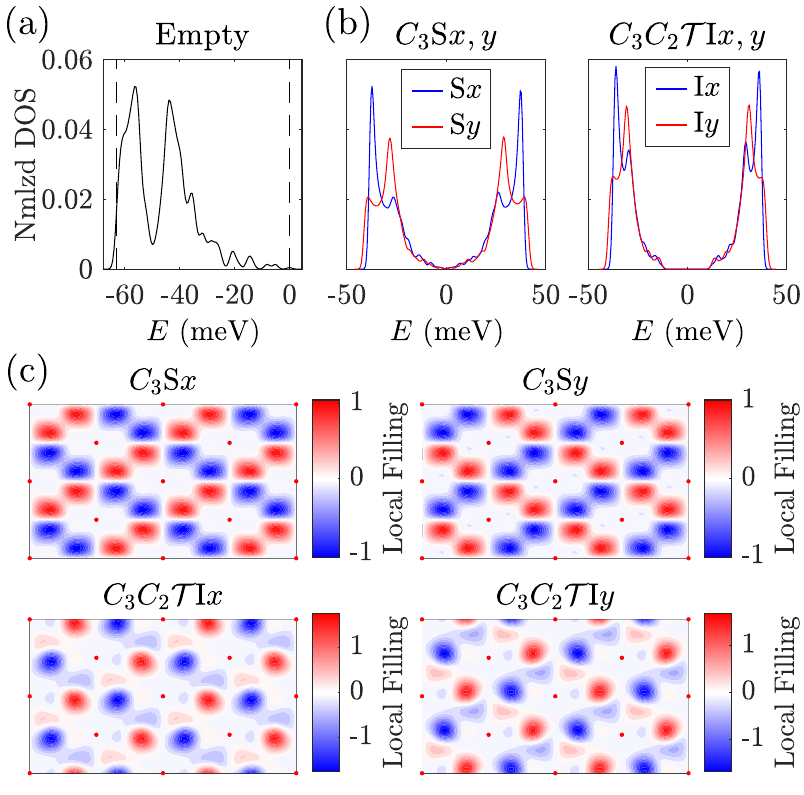}
	\caption{(a) Normalized DOS of the empty filling band structure. The two vertical dashed lines indicate the band bottom and top which is chosen to be at $E=0$. (b) Normalized DOS and (c) local filling fraction for the upper layer defined in (\ref{nur}) for the four potential ground states for strain parameter $\beta = \pm 4 \times 10^{-4}$. For positive (negative) beta: the two competing ground states are a $C_2 \T$-preserving semimetal $C_3$S$y$ ($C_3$S$x$) and a $C_2 \T$-breaking valley Hall insulator $C_3 C_2 \T$I$y$ ($C_3 C_2 \T$I$x$), both strongly breaking $C_3$. The red dots in (c) indicates the AA position. In (c) the breaking of $C_2$ and $C_3$ rotation symmetries are clearly visible.
	}
	\label{fig:DOSLocalFilling}
\end{figure}

Once $\beta$ becomes non-zero, the energy of the $C_2 \T$ breaking state is reduced relative to the VP/SP state. Furthermore, one of the two $C_3$-breaking states (depending on the sign of $\beta$) goes down in energy becoming more energetically favorable to the VP/SP state around $\beta = \pm 3 \times 10^{-4}$. For larger values of $\beta \gtrsim 4 \times 10^{-4}$, the energies of the insulating $C_2 \T$-breaking phase and the semimetallic $C_2 \T$-preserving phase approach the same value. This indicates that even very small explicit $C_3$-breaking picks out these two states as the main candidates for the ground state at CN. Assuming that such small explicit $C_3$ symmetry breaking exists in real samples due to strain, our analysis leads to the conclusion that the ground state of TBG at CN is either a $C_2 \T$-breaking insulator or a $C_2 \T$-symmetric semimetal. Both states strongly break $C_3$ symmetry and are very close in energy. It is worth noting that, in the presence of disorder, massless Dirac fermions may also emerge from the spatial domain walls of locally $C_2\mathcal{T}$ breaking insulating regions \cite{AliceaDisorder}.

\section{Energy competition in chiral limit}
The existence of several states close in energy can be explained by considering the simplified chiral model of Ref.~\cite{Tarnopolsky} where the Moir\'e intrasublattice hopping is switched off. In such limit, the band becomes strictly flat at charge neutrality in the absence of interaction and the model has an extra sublattice symmetry i.e. the single-particle Hamiltonian anticommutes with the sublattice operator $\sigma_z$. As a result, the flatband wavefunctions can be chosen to be eigenfunctions of $\sigma_z$ such that the wavefunction $\sigma_z u_{A/B,\tau,\bk} = \pm u_{A/B,\tau,\bk}$. In this limit, we can show (see Appendix for details) that the form factor matrix has the simple form
\beq
\Lambda_{\bq}(\bk) = F_\bq(\bk) e^{i\Phi_\bq(\bk) \sigma_z \tau_z}, 
\eeq
where $F_\bq(\bk)$ and $\Phi_\bq(\bk)$ denoting the magnitude and phase of the form factor for the flatband wavefunction in sublattice A and valley $K$.

Let us now compare the mean field energy for different phases. First, we notice that, even in the chiral limit where the non-interacting band dispersion vanishes, a sizable dispersion would be generated by the interaction \cite{KIVC}. It will be instructive, however, to start by ignoring the interaction generated band dispersion and focusing on the Hartree-Fock energy. At a fixed integer filling, the Hartree term does not play a role in the energy competition between phases so we can focus on the Fock term. We will find it convenient to write the projector $P(\bk)$ as
\beq
P(\bk) = \frac{1}{2}[1 + Q(\bk)], \qquad Q(\bk)^2 = 1, \quad \tr Q(\bk) = 0
\eeq
In terms of $Q$, the Fock energy is given by
\beq
E_F = - \frac{1}{8A} \sum_{\bk,\bq} V_{\bq} \tr Q(\bk) \Lambda^\dagger_{\bq}(\bk) Q(\bk + \bq) \Lambda_{\bq}(\bk). 
\label{EF}
\eeq
up to a constant. We note that $\langle A, B \rangle = \tr A B$ defines a positive definite inner product on the space of hermitian matrices. Using Cauchy-Schwarz inequality, we get
\beq
E_F \geq E_{F,\rm min} =  - \frac{1}{A} \sum_{\bk,\bq} V_{\bq} F^2_{\bq}(\bk).  
\label{EFboundVUP}
\eeq
This inequality is satisfied if and only if $Q(\bk + \bq)$ is parallel to $  \Lambda_{\bq}(\bk) Q(\bk) \Lambda^\dagger_{\bq}(\bk)$ for every $\bk$ and $\bq$ which implies
\beq
Q(\bk + \bq) = e^{i \Phi_\bq(\bk) \sigma_z \tau_z} Q(\bk) e^{-i \Phi_\bq(\bk) \sigma_z \tau_z}, 
\label{Pmin}
\eeq
A $\bk$-independent flavor (spin,valley) polarized state obviously satistfies this constraint since $\sigma_z \tau_z$ is diagonal in flavor, thereby saturating the Fock bound. This state can either be spin-polarized $Q = s_z$, valley polarized  $Q = \tau_z$ or spin-valley locked $Q = \tau_z s_z$. 

Next let us discuss $C_2\T$ breaking insulating solutions. A $C_2 \T$ breaking solution to Eq.~\ref{Pmin} is obtained by taking $Q_\bk = \sigma_z$ (since $C_2 \T$ is off diagonal in the sublattice index). The resulting state saturates the Fock bound and is characterized by a Chern number $\pm 1$ in the lower band. Its energy competition with the flavor-polarized states is settled by the effects of the single-particle term $h_\bk$ and $w_0$ which are neglected here. In our numerics, we found that these states are very close in energy for the realistic model with their energy difference only sensitive to the strain parameter $\beta$. Depending on its structure in flavor space, the $C_2 \T$-breaking insulator may correspond to one of four states: (i) $Q = s_0 \tau_0 \sigma_z$ breaks $C_2$ but not $\T$ and corresponds to a valley Hall state, (ii) $Q = s_0 \tau_z \sigma_z$ breaks $\T$ but not $C_2$ and corresponds to a quantum Hall state with Chern number $\pm 4$, (iii) $Q = s_z \tau_0 \sigma_z$ corresponds to a valley-spin-Hall state where the Chern number is invariant under flipping spin and valley, and (iv) $Q = s_z \tau_z \sigma_z$ corresponds to a spin-Hall state.

\begin{figure}
	\centering
	\includegraphics[width=1\linewidth]{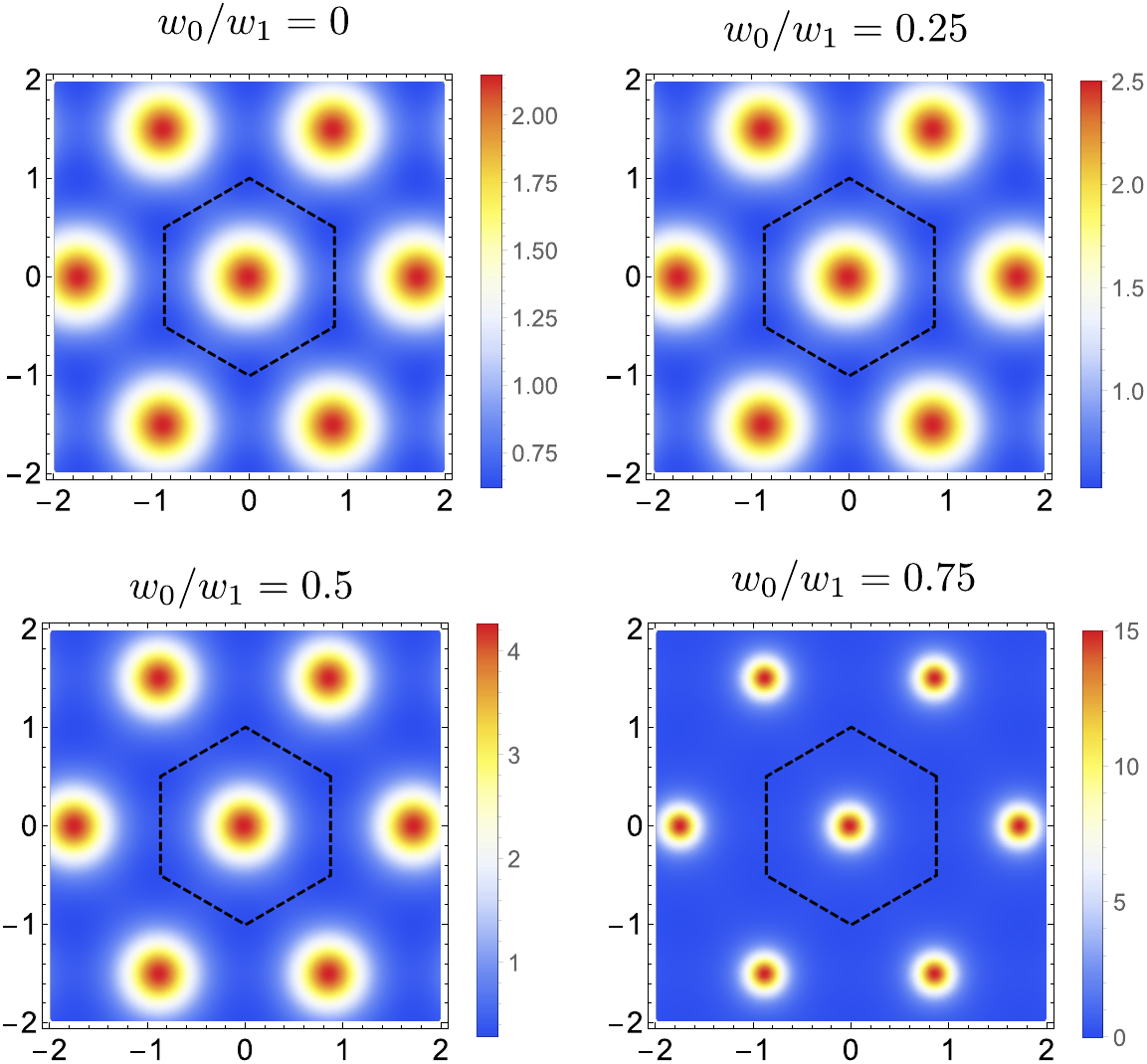}
	\caption{Berry curvature in the sublattice polarized basis for different values of the ratio $w_0/w_1$.
	}
	\label{fig:Fk}
\end{figure}

Finally, let us consider semimetallic states which break neither flavor nor $C_2 \T$ symmetry. These can be generally described (within each flavor) by the order parameter
\beq
Q_{\rm SM}(\bk) = \sigma_x e^{i \alpha_\bk \sigma_z \tau_z}. 
\label{QSM}
\eeq
Substituting in the Fock energy (\ref{EF}) yields
\beq
E_F = -\frac{1}{A} \sum_{\bk,\bq}
V_{\bq} F^2_{\bq}(\bk) \cos[ \alpha_\bk - \alpha_{\bk + \bq} + 2 \Phi_\bq(\bk)]. 
\label{EFSM}
\eeq 
To simplify further, we make the reasonable assumption that $F_\bq(\bk)$ decays quickly with the relative momentum $\bq$ which enables us to expand $\Phi_\bq(\bk)$ and $\alpha_{\bk + \bq}$ in $\bq$. Noting that $\Phi_\bq(\bk) = \bq \cdot \bA_\bk + O(\bq^2)$ with $\bA_\bk$ the Berry connection $-i \langle u_{A,K,\bk}|\nabla_\bk|u_{A,K,\bk}\rangle$, we get
\beq
    E_F = E_{F,\rm min} + \frac{1}{2A} \sum_{\bk,\bq} V_\bq \bq^2 F^2_{\bq}(\bk) (\nabla_\bk \alpha_\bk - 2\bA_\bk)^2. 
    \label{EFSM2}
\eeq
The second term in the energy is non-negative and implies that the semimetal does not generally satisfy the bound. In fact, this term has the form of the energy of a superconductor in a magnetic field in momentum space. Due to the non-trivial Chern number of the sublattice-polarized bands \cite{Tarnopolsky, KIVC}, the Bloch wavefunctions cannot be chosen to be simultanuously smooth and periodic over the Brillouin zone. If we choose them to be smooth, the Berry connection $\bA_\bk$ will be smooth but the phase $\alpha_\bk$ will necessarily wind by $4\pi$ around the Brillouin zone, leading to at least two vortices in the Brillouin zone where the projector is undefined due to the vanishing of the gap. These will yield a finite contribution to the second term which implies that the semimetal never satisfies the Fock bound in the ideal limit \footnote{Notice that alternatively, if we choose a periodic gauge, $\alpha_\bk$ can be made continuous but the Berry connection will be singular leading also to finite energy contribution}. The location of the vortices which minimize this term is determined by the competition between the weak (logarithmic) repulsion between the vortices in 2D and the tendency of the vortices to sit wherever the magnetic field corresponding to $\bA$, i.e. the Berry curvature, is maximal. As shown in Fig.~\ref{fig:Fk}, in the chiral limit $w_0 = 0$, the Berry curvature is relatively uniform so that vortices prefer to sit far apart at the K and K$'$ point, thereby preserving $C_3$ symmetry. However, for finite $w_0/w_1$, the Berry curvature (in the basis which diagonalizes the sublattice operator $\sigma_z$ \cite{KIVC}) becomes strongly peaked at the $\Gamma$ point and for the realistic parameter $w_0/w_1 \approx 0.75$, the vortices prefer to sit close to the $\Gamma$ point. This explains the energetic advantage of $C_3$ symmetry breaking in the realistic parameter regime considered in the numerics. 

Finally, let us consider the effect of the dispersion (which is mostly generated by the interaction). Using $C_2 \T$, valley and particle-hole symmetries, the form of the dispersion can be reduced to \cite{KIVC}
\beq
h_0(\bk) = \sigma_x f(\bk) e^{i \phi(\bk) \sigma_z \tau_z}
\eeq
where $f(\bk)$ is a positive real function which vanishes at the two Dirac points and $\phi(\bk)$ winds by $2\pi$ around each of the Dirac points. These points lie at the Moir\'e $K_M$ and $K'_M$ for $\beta = 0$ but move away for finite $\beta$. The energy contribution of $h_0(\bk)$ for a state described by the matrix $Q$ is
\beq
E_h[Q] = \frac{1}{2} \sum_\bk \tr h_0(\bk) Q(\bk)
\label{EhQ}
\eeq
We notice that this term favors the semimetal order parameter with $Q$ given by (\ref{QSM}) but vanishes for all the order parameters which satisfy the Fock bound. This might suggest none of these states benefit energetically from dispersion but this turns out not to be true as we will show below.

We start by writing the total energy functional in terms of $Q$ (ignoring the Hartree term which does not play a role in the competition between phases)
\beq
E[Q] = E_h[Q] + E_F[Q]
\label{EQ}
\eeq
with $E_h[Q]$ given by (\ref{EhQ}) and $E_F[Q]$ given by (\ref{EF}). The Hartree-Fock self consistency equation is taken by setting the variation of $E[Q]$ relative to $Q$ to zero (subject to the constraint $Q^2 = 1$) which leads to the equation
\beq
[Q(\bk), \frac{1}{2} h_0(\bk) - \frac{1}{4A} \sum_\bq V_\bq(\bk) \Lambda_\bq(\bk)^* Q(\bk + \bq) \Lambda_\bq(\bk)^T] = 0
\label{SC}
\eeq
From this equation, we see that the flavor polarized states which are characterized by $[Q, h_0] = 0$ remain solution of the self-consistency equation for non-zero $h_0$ whereas the $C_2 \T$ breaking insulators which are characterized by $\{Q, h_0\} = 0$ are no longer solution to the self-consistency equation. To write a $C_2 \T$ breaking solution, we write
\beq
Q(\bk) = Q_{C_2\T \rm I}(\bk) \cos \theta(\bk) + Q_{\rm SM}(\bk) \sin \theta(\bk)
\label{QC2TSM}
\eeq
which satisfies the condition $Q^2 = 1$ due to the anticommutation of $Q_{C_2\T \rm I}$ and $Q_{\rm SM}$. The $\bk$ dependence of $\theta$ can be determined from the self-consistency condition (\ref{SC}). In the following,  we will assume for simplicity that $\theta$ is $\bk$ independent and choose it to minimize the the total energy (\ref{EQ}) leading to
\beq
|\sin \theta| = \frac{E_h[Q_{\rm SM}]}{2(E_F[Q_{\rm SM}] - E_{F,\rm min})}
\eeq
where we used the fact that $Q_{C_2\T \rm I}$ saturates the Fock bound. We notice that the denominator is the same as the second term on the right hand side of (\ref{EFSM2}) which is guaranteed to be finite and roughly of the order of the interaction scale. This implies that $\theta$ is small whenever the interaction dominates the dispersion. In this case, the order parameter described by (\ref{QC2TSM}) is very close to the pure insulator order parameter. The corresponding energy is given by
\beq
E[Q] = E_{F,\rm min} - \frac{E_h[Q_{\rm SM}]^2}{4(E_F[Q_{\rm SM}] - E_{F,\rm min})}
\eeq
Thus, the $C_2\T$-breaking insulators actually benefit from the dispersion in quadratic order by developing a small component proportional to the semimetal order parameter. This explains why their energy is decreased in response to explicit $C_3$-breaking which energetically favors the semimetal (thus increasing $|E_h[Q_{\rm SM}]|$). It also explains their high $C_3$-breaking susceptibility which mainly arises from their $Q_{\rm SM}$ component. 

\section{Consequences for Experiment}
The possibility of $C_3$ breaking at CN is consistent with several recent reports \cite{CaltechSTM, ColombiaSTM, RutgersSTM, PrincetonSTM} which observed direct evidence of $C_3$ breaking in STM measurements. To check the compatibilty of these measurements with our mean-field solutions, we compute the DOS for the four possible $C_3$-breaking states (arising for positive or negative values of $\beta$) in Fig.~\ref{fig:DOSLocalFilling}b. We see that in all cases the global DOS consists of two broad peaks (which are sometimes further split into two) separated by about 60-80 meV, in agreement with the STM measurements \cite{ColombiaSTM, CaltechSTM}. The DOS alone however, is insufficient to distinguish the insulating and semi-metallic state, since both have very low DOS close to zero energy. One way to distinguish the two is to compute the local filling fraction defined as \cite{RutgersSTM}
\beq
\nu(\br) = 8 \left(\frac{\rho_{\rm LB}(\br)}{\rho_{\rm LB}(\br) + \rho_{\rm UB}(\br)} - \frac{1}{2} \right)\in[-4,4], 
\label{nur}
\eeq
where $\rho_{\rm LB/UB}(\br)$ denote the integrated local DOS from the upper layer for the lower/upper band. $\nu(\br)$, shown in Fig.~\ref{fig:DOSLocalFilling}, exhibits clear $C_3$-symmetry breaking pattern for all the four phases. The patterns for the $C_3$S and $C_3C_2\T$I QVH phases can be distinguished by $C_2$ symmetry which is visibly present in the former but absent in the latter (the pattern of the other $C_3C_2\T$I is obtained by symmetrizing the QVH result relative to $C_2$ with the result being very similar to the one for the $C_3$S breaking states up to an overall factor). 
The $C_3$S pattern is qualitatively similar to that measured in Ref.~\cite{RutgersSTM}, with the density vanishing at the Kagom\'e lattice site lying on the mirror plane and changing sign twice around it, while having non-vanishing magnitude with opposite signs on the other two Kagom\'e lattice sites. Such structure is generic for any $C_3$-breaking phase which preserves mirror and particle-hole symmetries \footnote{The BM Hamiltonian has an approximate particle-hole symmetry at small angles} and a combination of $C_2$ and some local symmetry. \cite{Hejazi19}   

\section{Conclusion}
In conclusion, we have performed a momentum-space self-consistent Hartree-Fock analysis to uncover the nature of the symmetry-broken phase in twisted bilayer graphene at charge neutrality. In addition to insulating states corresponding to spin, valley or sublattice polarization, we found two $C_3$-breaking $C_2\T$-symmetric semimetallic solutions. Our main finding is that the existence of very small explicit $C_3$-breaking energetically favors one of these $C_2 \T$-symmetric metallic state together with $C_2 \T$-breaking 
insulating states. Both sets of states have similar energy within HF, strongly break $C_3$ symmetry and are consistent with the density of states measured in STM experiments. They can be experimentally distinguished in transport measurements or by comparing space-resolved local filling fractions in STM. We propose these two states as candidates for the insulating and conducting states observed in different experiments at the CNP and suggest that the competition between the two is settled by small details that are likely sample-dependent. 

\acknowledgements
We thank Eva Andrei, Allan MacDonald, Adrian Po, A. Thomson and M. Xie for helpful discussions. 
S. L., E. K., J. Y. L. and A. V. were supported by a Simons Investigator Fellowship and by NSF-DMR 1411343. E. K. was supported by the German National Academy of Sciences Leopoldina through grant LPDS 2018-02 Leopoldina fellowship.

\appendix
\begin{widetext}
	\section{Single-particle physics: Bistritzer-Macdonald model}
	
	Our starting point is the Bistritzer-MacDonald (BM) model of the TBG band structure \cite{MacDonald2011}, which we now briefly review. We begin with two layers of perfectly aligned (AA stacking) graphene sheets extended along the $xy$ plane, and we choose the frame orientation such that the $y$-axis is parallel to some of the honeycomb lattice bonds. Now we choose an arbitrary atomic site and twist the top and bottom layers around that site by the counterclockwise angles $\theta/2$ and $-\theta/2$ (say $\theta>0$), respectively. When $\theta$ is very small, the lattice form a Moir\'e pattern with very large translation vectors; correspondingly, the Moir\'e Brillouin zone (MBZ) is very small compared to the monolayer graphene Brillouin zone (BZ), as illustrated in Fig.~\ref{rSpace&kSpace}. In this case, coupling between the two valleys can be neglected. If we focus on one of the two valleys, say $K$, then the effective Hamiltonian is given by: 
	\begin{align}
	\H_+= \sum_l \sum_{\bk} \tilde f^\dagger_{l}(\bk) h_{\bk}(l\theta/2) \tilde f_{l}(\bk) + \left(\sum_{\bk}\sum_{i=1}^3 \tilde f^\dagger_{t} (\bk+\bq_i) T_i \tilde f_{b}(\bk) + h.c.\right). \end{align}
	Here, $l=t/b\simeq \pm 1$ is the layer index, and $\tilde f_l(\bk)$ is the $K$-valley electron originated from layer $l$. The sublattice index $\sigma$ is suppressed, thus each $\tilde f_l(\bk)$ operator is in fact a two-component column vector. $h_{\bk}(\theta)$ is the monolayer graphene $K$-valley Hamiltonian with twist angle $\theta$: 
	\begin{align}
	h_{\bk}(\theta)=\hbar v_F
	\begin{pmatrix}
	0 & (k_x-i k_y) e^{i\theta}\\
	(k_x+i k_y) e^{-i \theta} & 0
	\end{pmatrix}, 
	\end{align}
	where $v_F=9.1\times 10^5$~m/s is the Fermi velocity. Let $K_l$ be the $K$-vector of layer $l$, then $\bq_1$ is defined as $K_b-K_t$. $\bq_2=O_3\bq_1$ is the counterclockwise $120^\circ$ rotation of $\bq_1$, and $\bq_3=O_3\bq_2$. Finally, the three matrices $T_i$ are given by 
	\begin{align}
	T_1=(1 + \beta)
	\begin{pmatrix}
	w_0 & w_1\\
	w_1 & w_0
	\end{pmatrix}, \quad
	T_2=
	\begin{pmatrix}
	w_0 & w_1 e^{-2\pi i/3}\\
	w_1 e^{2\pi i/3} & w_0
	\end{pmatrix}, \quad
	T_3=
	\begin{pmatrix}
	w_0 & w_1 e^{2\pi i/3}\\
	w_1 e^{-2\pi i/3} & w_0
	\end{pmatrix}, 	
	\label{T123}
	\end{align}
	where we introduced an explicit $C_3$-breaking parameter $\beta$. We take $w_1=110$~meV and $w_0=82.5$~meV. The difference between $w_0$ and $w_1$ reflects the effect of lattice relaxation. Note that the argument of $\tilde f_l(\bk)$ is measured from $K_l$, i.e. the monolayer UV momentum associated to $\tilde f_l(\bk)$ is in fact $K_l+\bk$. It is convenient to also choose a common momentum reference point for the two layers. For example, we can define $f_t(\bk)=\tilde f_t(\bk-\bq_3)$ and $f_b(\bk)=\tilde f_b(\bk+\bq_2)$, such that the arguments for both $f_t$ and $f_b$ are measured from $K_t-\bq_3$, indicated as $\gamma$ in the right panel of Fig.~\ref{rSpace&kSpace}. 
	
	One intuitive way of thinking about this effective Hamiltonian is to imagine a honeycomb lattice of Dirac points in the momentum space, as shown in Fig.~\ref{MagicAngleBS&DiracPointLattice}b, where the two sublattices correspond to the two layers. A Dirac point at momentum $\bq$ contributes a diagonal block $h_{\bk-\bq}(\pm\theta/2)$ to the Hamiltonian for the MBZ momentum $\bk$, where the sign is determined by the sublattice that $\bq$ belongs to. The off-diagonal blocks $T_i$ are nothing but the nearest-neighbor couplings of these Dirac points. 
	
	\begin{figure}[h]
		\centering
		\includegraphics[width=0.5\linewidth]{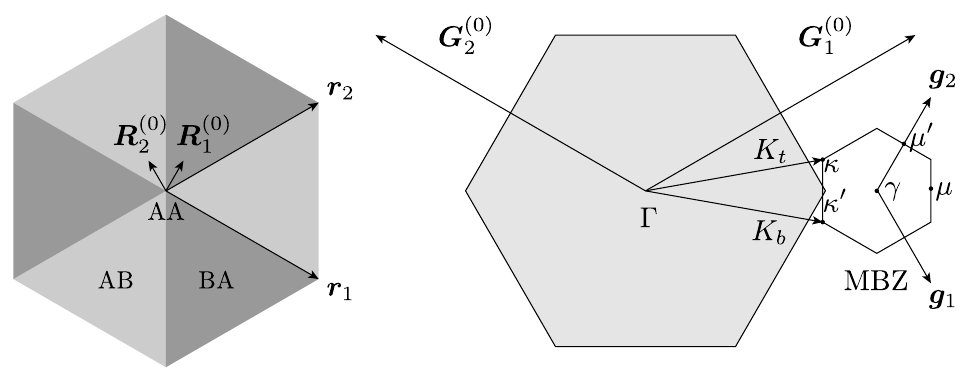}
		\caption{Real space (left) and momentum space (right) structures for both monolayer and Moir\'e lattices of twisted bilayer graphene. In the real space (left) panel, we also show the underlying Moir\'e pattern structure. } 
		\label{rSpace&kSpace}
	\end{figure}
	
	\begin{figure}[h]
		\centering
		\includegraphics[width=0.5\linewidth]{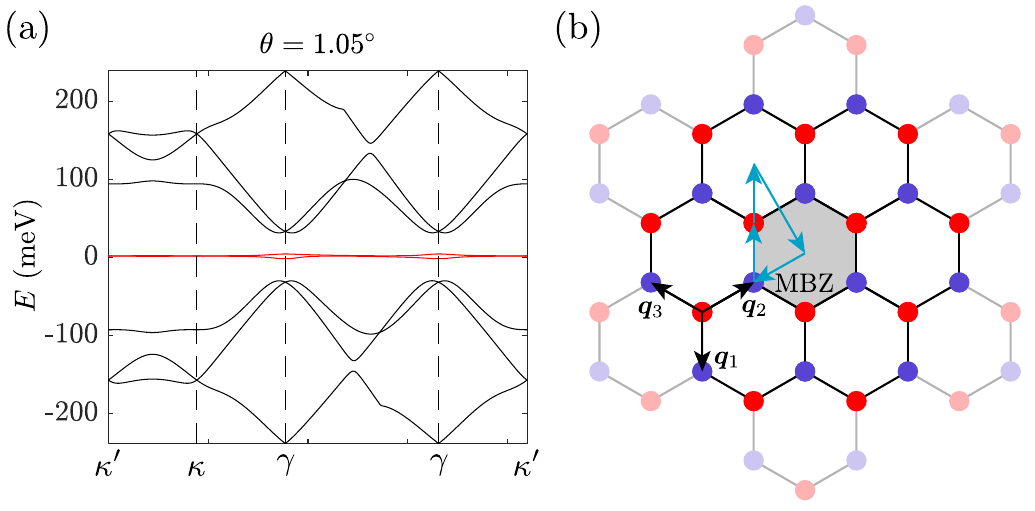}
		\caption{(a) Band structure of magic-angle TBG obtained from the BM model. The sampling path is shown by the cyan arrows in the right panel. (b) Dirac point lattice at the MBZ. }
		\label{MagicAngleBS&DiracPointLattice}
	\end{figure}
	
	When the twist angle is near the magic angle $\theta=1.05^\circ$, two isolated flat bands per spin and valley appear near the charge neutrality (CN) Fermi energy, shown in Fig.~\ref{MagicAngleBS&DiracPointLattice}a. These two bands are the focus of the current work. 
	
	The single particle Hamiltonian within each valley $\H_\pm$ is invariant under the following symmetries
	\beq
	C_3 \tilde f_{\bk} C_3^{-1} = e^{-\frac{2\pi}{3} i \tau_z \sigma_z} \tilde f_{C_3 \bk}, \quad (C_2 \T) \tilde f_\bk (C_2 \T)^{-1} = \sigma_x \tilde f_\bk, \qquad \M_y \tilde f_\bk \M_y^{-1} = \sigma_x \mu_x \tilde f_{M_y \bk},
	\eeq
	In addition, the two valleys are related by time-reversal symmetry given by
	\beq
	\T \tilde f_\bk \T^{-1} = \tau_x \tilde f_{-\bk}.
	\eeq
	Here, $\boldsymbol\sigma, \boldsymbol\tau$ and $\boldsymbol \mu$ denote the Pauli matrices in sublattice, valley and layer spaces, respectively.
	
	\section{Projecting the interaction onto the flat bands}
	In the following, we derive the form of the interaction when projecting onto the two flat bands. Since these two bands have a Wannier obstruction, we can only write such projected interaction in $k$-space. Let $c^\dagger_{\alpha}(\bk)$ be the creation operator for the energy eigenstate in the band structure with internal flavor $\mu$ and band index $n$, where $\mu=(\tau,s)$ is a collective index including both valley $\tau=\pm$ and spin $s=\uparrow/\downarrow$, and $n=1,2$ represents the lower and upper bands, respectively. Also let $f^\dagger_{\mu,I}(\bq)$ be the ``elementary'' continuous fermion with monolayer momentum $\bq$, flavor $\mu=(\tau,s)$ and $I=(l,\sigma)$ representing layer and sublattice, then $c^\dagger$ and $f^\dagger$ are related to each other by the $k$-space wave functions as follows: 
	\begin{align}
	c^\dagger_{\mu,n}(\bk)=\sum_{\bG,I} u_{\tau,n;\bG,I}(\bk) f^\dagger_{\mu,I}(\bk + \bG), 
	\end{align}
	where $\bG$ is a Moir\'e reciprocal lattice vector. In the above expression, we are already using the fact that the wave functions are spin-independent. Once we choose a gauge of $u_{\tau,n;\bG,I}(\bk)$ for all $\bk$ in some MBZ, $c^\dagger(\bk)$ are defined in terms of the $f^\dagger(\bq)$ for those $\bk$, and whenever necessary, we define $c^\dagger(\bk+\bG) = c^\dagger(\bk)$ for any reciprical lattice vector $\bG$, which is equivalent to defining $u_{\tau,n;\bG,I}(\bk+\bG_0)=u_{\tau,n;\bG + \bG_0,I}(\bk)$. Note that the momentum argument for $f^\dagger$ is unconstrained since we are using the continuum theory for monolayers of graphene. We choose the normalization $\{f_{\mu,I}(\bq),f^\dagger_{\mu',I'}(\bq')\}=\delta_{\mu\mu'} \delta_{II'} \delta_{\bq\bq'}$ (suppose the system size is finite), and $\inner{u_{\tau,n}(\bk)}{u_{\tau',n'}(\bk)}:=\sum_{\bG,I} u^*_{\tau,n;\bG,I}(\bk)u_{\tau',n';\bG,I}(\bk)=\delta_{\tau \tau'}\delta_{nn'}$, which imply $\{ c_{\mu,n}(\bk),c^\dagger_{\mu',n'}(\bk')\}=\delta_{\mu\mu'}\delta_{nn'}\delta_{\bk\bk'}$ when $\bk,\bk'$ are confined in the MBZ. For the purpose of projecting the interaction into these two bands, it is convenient to introduce the form factor notation: 
	\begin{align}
	\lambda_{mn,\tau;\bq}(\bk) := \inner{ u_{\tau,m}(\bk) } { u_{\tau,n}(\bk + \bq) } 
	\end{align}
	where $\bq$ is not restricted to the first Brillouin zone. The form factors satisfy 
	\begin{align}
	\lambda_{mn,\tau;\bq}(\bk)=\lambda_{nm, \tau;-\bq}(\bk + \bq)^*
	\end{align}
	just from the definition, and also has the property
	\begin{align}
	\lambda_{mn, \tau; \bq}(\bk) = \lambda_{mn, -\tau; -\bq}(-\bk)^*
	\end{align}
	due to the time-reversal symmetry. 
	
	The interaction Hamiltonian is given by
	\beq
	\H_{\rm int} = \frac{1}{2A} \sum_{\sigma, \sigma', \tau, \tau'} \sum_{\bq} V(\bq) :\rho_{\sigma, \tau,\bq}  \rho_{\sigma', \tau',-\bq}:,
	\eeq
	where $A$ is the total area of the system and $V(\bq)$ is the momentum space interaction potential, related to the real-space one by $V(\bq):= \int d^2 \br V(\br) e^{-i \bq \cdot \br}$. Depending on the number of gates, $V(\bq)$ takes the following form in the SI units: 
	\begin{align}
	V(\bq)= \frac{e^2}{2 \epsilon\epsilon_0 q}
	\begin{cases}
	(1-e^{-2q d_s}), &(\text{single-gate})\\
	\tanh(q d_s), &(\text{dual-gate})
	\end{cases}
	\end{align}
	where the screening length $d_s$ is nothing but the distance from the graphene plane to the gate(s).  Projecting onto the two narrow bands, this Hamiltonian has the form
	\begin{multline}
	\H_{\rm int} = \frac{1}{2A} \sum_{\sigma, \sigma', \tau, \tau'} \sum_{\bq,n_1, n_2, n_3, n_4} \sum_{\bk_1,\bk_2 \in \rm BZ} \lambda_{n_1, n_2;\tau,\bq}(\bk_1) V(\bq) \lambda^*_{n_4, n_3;\tau',\bq}(\bk_2) \\ \times c^\dagger_{n_1, \sigma,\tau}(\bk_1)  c^\dagger_{n_3, \sigma',\tau'}(\bk_2+\bq) c_{n_4, \sigma',\tau'}(\bk_2) c_{n_2, \sigma,\tau}(\bk_1 + \bq). 
	\end{multline}

	\section{Hartree-Fock analysis in the chiral limit}
	The chiral limit of the BM model is obtained by switching off the $w_0$ term in (\ref{T123}) \cite{Tarnopolsky}. In this limit, the bands become exactly flat at the magic angle and the eigenstates of the Hamiltonian have a simple form similar to the Landau levels on a torus 
	In the chiral limit, the single particle Hamiltonian anticommutes with the chiral (sublattice) symmetry operator given by $\Gamma = \sigma_z$. In addition, it is invariant under the particle-hole symmetry $\P \tilde f_\bk \P^{-1} = i\sigma_x \mu_y \tilde f^\dagger_{-\bk}$ (modulo a small basis rotation gauging away the $\theta$ dependence). This means that we can choose the wavefunctions for the flat bands to be eigenfunctions of the sublattice operator $\sigma_z$ i.e. completely sublattice polarized \cite{KIVC}. The wavefunctions can then be labelled by their sublattice index $\sigma = $A/B. This means that the sublattice off-diagonal components of the form factor vanishes, i.e. $\lambda_{AB;\tau,\bG}(\bk, \bk') = \lambda_{BA;\tau,\bG}(\bk, \bk') = 0$.  Furthermore, the action of $C_2 \T$ is given by
	\beq
	C_2 \T u_{A,\tau,\bk} = e^{i \phi_{A,\tau,\bk}} u^*_{B,\tau,\bk}, \qquad C_2 \T u_{B,\tau,\bk} = e^{i \phi_{B,\tau,\bk}} u^*_{A,\bk}
	\eeq
	which implies that $\lambda_{AA;\tau,\bq}(\bk) = \lambda_{BB;\tau,\bq}(\bk)^*$. Finally, we can use $\P \T$ symmetry to restrict the form factors further by noting that it exchanges positive and negative energy eigenstates in opposite valleys and sublattices
	\beq
	\P \T u_{A,\tau,\bk} = e^{i \eta_{A,\tau,\bk}} u_{B,-\tau,\bk}, \qquad \P \T u_{B,\tau,\bk} = e^{-i \eta_{B,\tau,\bk}} u_{A,-\tau,\bk}
	\eeq
	which implies that $\lambda_{A/B,\tau,\bq}(\bk) = \lambda_{B/A,-\tau,\bq}(\bk)$. Summarizing these conditions, we can write
	\beq
	\lambda_{\sigma \sigma';\tau,\bq}(\bk) = \delta_{\sigma \sigma'} F_\bq(\bk) e^{i\Phi_\bq(\bk) \sigma \tau}
	\eeq
	where $F_\bq(\bk) = |\lambda_{AA,+,\bq}(\bk)|$ and $\phi_\bq(\bk) = \arg \lambda_{AA,+,\bq}(\bk)$
	
	We now investigate the Hartree-Fock solutions by looking for the minima of the Hartree-Fock energy. The Hartree-Fock energy is defined in terms of the order parameter
	\beq
	P_{\alpha \beta}(\bk) = \langle c^\dagger_\alpha(\bk) c_\beta(\bk) \rangle
	\eeq
	where $\alpha, \beta$ range over spin, valley and band indices. For an insulator or a semimetal, the number of filled states is $\bk$ independent and equal 4. This means that the order parameter $P_\bk$ is a projector satisfying
	\beq
	P(\bk)^2 = P(\bk) = P^\dagger(\bk), \qquad \tr P(\bk) = 4
	\eeq
	
	The Hartree-Fock energy can then be written as (using properties of the chiral limit)
	\begin{gather}
	E_{\rm HF} = E_H + E_F,  \\
	E_H =  \frac{1}{2A} \sum_{\bG,\bk,\bk'} V_\bG \tr P(\bk) \Lambda^\dagger_\bG(\bk) \tr P(\bk') \Lambda_\bG(\bk'), \\
	E_F = - \frac{1}{2A} \sum_{\bq,\bk} V_{\bq} \tr P(\bk) \Lambda^\dagger_{\bq}(\bk) P(\bk + \bq) \Lambda_{\bq}(\bk) 
	\end{gather}
	where we defined the form factor matrix $\Lambda_\bq(\bk)$ in terms of the combined index $\alpha = (s, \tau, n)$ for spin, valley, and band as
	\beq
	\Lambda_\bq(\bk) = F_\bq(\bk) e^{i \Phi_\bq(\bk) \sigma_z \tau_z}
	\eeq
	The form factors decay in the separation $\bq$ with a characteristic scale which is typically smaller than the size of the Brillouin zone. This means that the sums in the Hartree and Fock terms are dominated by the $\bG = 0$ term. For the Hartree term, this equals $4 V(0) N$ and is independent of the order parameter $P(\bk)$. Thus, the Hartree term has little effect on the competition between different phases and and will be neglected in the following.  
	
	We now consider possible types of symmetry-broken orders at charge neutrality. The order parameter $P(\bk)$ can generally be written as 
	\beq
	P(\bk) = \frac{1}{2}[1 + Q(\bk)], \qquad \tr Q(\bk) = 0
	\eeq
	Since the Fock term is the largest contribution to the mean-field energy, let us now neglect the Hartree term as well as the single-particle term $h_0$. We note that $\langle A, B \rangle = \tr A B$ defines a positive definite inner product on the space of hermitian matrices. Using Cauchy-Schwarz inequality, we get
	\beq
	E_F \geq - \frac{1}{2A} \sum_{\bk,\bq} V_{\bq} \sqrt{\tr P(\bk) \tr [\Lambda^\dagger_{\bq}(\bk) P(\bk') \Lambda_{\bq}(\bk)]^2} = - \frac{2}{A} \sum_{\bk,\bq} V_{\bq} F^2_{\bq}(\bk) 
	\label{EFboundVUP}
	\eeq
	This inequality is satisfied if and only if $P(\bk + \bq)$ is parallel to $  \Lambda_{\bq}(\bk) P(\bk) \Lambda^\dagger_{\bq}(\bk)$ for every $\bk$, $\bq$. This means
	\beq
	P(\bk+\bq) = e^{i \Phi_\bq(\bk) \sigma_z \tau_z} P(\bk) e^{-i \Phi_\bq(\bk) \sigma_z \tau_z} \quad \Rightarrow \quad Q(\bk + \bq) = e^{i \Phi_\bq(\bk) \sigma_z \tau_z} Q(\bk) e^{-i \Phi_\bq(\bk) \sigma_z \tau_z}
	\eeq
	
	\subsection{Intervalley coherent states}
	Let us now briefly discuss states which break U(1) valley symmetry. These states were argued to be energetically unfavorable in the context of twisted bilayer graphene with an aligned hBN substrate \cite{Bultinck19} and other Moir\'e materials which lack $C_2$ symmetry \cite{Zhang2018, Lee19}. Here, we will show that a similar argument \emph{fails} \footnote{We thank Mike Zaletel for discussions on this point} in $C_2$-symmetric twisted bilayer graphene due to the extra particle-hole symmetry $\P$ which is exact in the chiral limit. A more detailed discussion of such phases is provided in Ref.~\cite{KIVC}. To clarify the importance of $\P$, we will assume that the symmetry is not present in which case, the form factors in the two opposite valleys at the same momentum $\bk$ are not related, i.e. $\lambda_{A/B,+,\bq}(\bk) \neq \lambda_{B/A,-,\bq}(\bk)$. Furthermore, we will assume unbroken time-reversal symmetry. The case of time-reversal symmetry breaking can be addressed similarly. The projector for an IVC state can be split into a diagonal and off-diagonal component in valley space
	\beq
	P(\bk) = P_d(\bk) + P_o(\bk), \qquad P_d(\bk)^2 + P_o(\bk)^2 = P_d(\bk), 
	\eeq
	In terms of valley resolved blocks of $P(\bk)$, i.e.
	\beq
	P=
	\begin{pmatrix}
		P_+ & P_{12}\\
		P_{21} & P_-
	\end{pmatrix}, 
	\eeq
	the second condition can be written as
	\beq
	P_+^2+P_{12}P_{21}=P_+, \qquad P_-^2+P_{21}P_{12}=P_-. 
	\eeq
	Since the form factors are diagonal in valley space, the Fock energy can be written as a sum of a term with only diagonal part and one with only off-diagonal parts as
	\begin{align}
	E_F &= - \frac{1}{2A} \sum_{\bk,\bq} V_{\bq} \tr[ P_d(\bk) \Lambda^\dagger_{\bq}(\bk) P_d(\bk + \bq)\Lambda_{\bq}(\bk) + P_o(\bk) \Lambda^\dagger_{\bq}(\bk) P_o(\bk + \bq) \Lambda_{\bq}(\bk)] \nonumber \\
	&\geq - \frac{1}{2A} \sum_{\bk,\bq} V_{\bq} [|\lambda_{A,+,\bq}(\bk)|^2 \sqrt{\tr P_+(\bk)^2 \tr P_+(\bk + \bq)^2} + |\lambda_{A,-,\bq}(\bk)|^2 \sqrt{\tr P_-(\bk)^2 \tr P_-(\bk + \bq)^2}  \nonumber \\
	& \qquad \qquad + |\lambda_{A,+,\bq}(\bk)| |\lambda_{A,-,\bq}(\bk)| \sqrt{\tr P_o(\bk)^2 \tr P_o(\bk + \bq)^2}] \nonumber \\
	&\geq - \frac{1}{2A} \sum_{\bk,\bq} V_{\bq} [|\lambda_{A,+,\bq}(\bk)|^2 \tr P_+(\bk)^2 + |\lambda_{A,-,\bq}(\bk)|^2\tr P_-(\bk)^2  + |\lambda_{A,+,\bq}(\bk)| |\lambda_{A,-,\bq}(\bk)| \tr P_o(\bk)^2] \nonumber \\
	&= - \frac{1}{2A} \sum_{\bk,\bq} V_{\bq} [|\lambda_{A,+,\bq}(\bk)|^2 \tr (P_+(\bk)-P_{12}(\bk)P_{21}(\bk)) + |\lambda_{A,+,\bq}(\bk)|^2 \tr (P_-(\bk)-P_{21}(\bk)P_{12}(\bk)) \nonumber \\
	& \qquad \qquad + 2 |\lambda_{A,+,\bq}(\bk)| |\lambda_{A,-,\bq}(\bk)| \tr(P_{12}(\bk)P_{21}(\bk))] \nonumber \\
	&= - \frac{1}{A} \sum_{\bk,\bq} V_{\bq} (|\lambda_{A,+,\bq}(\bk)|^2 + |\lambda_{A,-,\bq}(\bk)|^2) + \frac{1}{4A} \sum_{\bk,\bq} V_{\bq} (|\lambda_{A,+,\bq}(\bk)| - |\lambda_{A,-,\bq}(\bk)|)^2 \tr P_o(\bk)^2, 
	\label{EFboundIVC}
	\end{align}
	Here, we used the Cauchy-Shwarz inequality to go from the first to the second line. We then used the geometric-arithmetic mean inequality $\sqrt{x y} \leq \frac{x + y}{2}$ to go from the second to the third. We now see that whenever the second term in the last line is non-zero, we would conclude that the Fock energy for the IVC is larger than the energy bound for the valley polarized or unpolarized phases by an amount which is proportional to the valley off-diagonal part of the order parameter. This energy difference in (\ref{EFboundIVC}) is generally not expected to be small unless there is a symmetry relating the same mometa in the two valleys (or equivalently a symmetry relating opposite momenta in the same valley). This is precisely the reason why the existence of particle-hole $\P$ symmetry forces the second term to vanish which makes the bound (\ref{EFboundIVC}), while correct, inconclusive to rule out IVC states. 
	
\end{widetext}
\bibliography{main.bib}
\end{document}